\documentclass[twocolumn,english,aps,prl,showpacs,superscriptaddress,floats,amsmath,amssymb,floatfix,nobalancelastpage]{revtex4}
\usepackage[T1]{fontenc}
\usepackage[latin9]{inputenc}
\usepackage{color}
\usepackage{amsmath}
\usepackage{amssymb}
\usepackage{graphicx}
\usepackage{esint}

\makeatletter
\@ifundefined{textcolor}{}
{%
 \definecolor{BLACK}{gray}{0}
 \definecolor{WHITE}{gray}{1}
 \definecolor{RED}{rgb}{1,0,0}
 \definecolor{GREEN}{rgb}{0,.7,0}
 \definecolor{BLUE}{rgb}{0,0,1}
 \definecolor{CYAN}{cmyk}{1,0,0,0}
 \definecolor{MAGENTA}{cmyk}{0,1,0,0}
 \definecolor{YELLOW}{cmyk}{0,0,1,0}
}

\usepackage{array}
\usepackage{babel}\@ifundefined{definecolor}
 {\usepackage{color}}{}

\allowdisplaybreaks

\newcommand\varpm{\mathbin{\vcenter{\hbox{%
  \oalign{\hfil$\scriptstyle+$\hfil\cr
          \noalign{\kern-.3ex}
          $\scriptscriptstyle({-})$\cr}%
}}}}
\newcommand\varmp{\mathbin{\vcenter{\hbox{%
  \oalign{$\scriptstyle({+})$\cr
          \noalign{\kern-.3ex}
          \hfil$\scriptscriptstyle-$\hfil\cr}%
}}}}
\makeatother

\usepackage{babel}
\begin{document}

\title{Thermal Transport in a one-dimensional Z$_{2}$ Spin Liquid }

\author{Alexandros Metavitsiadis}

\email{a.metavitsiadis@tu-bs.de}

\selectlanguage{english}%

\affiliation{Institute for Theoretical Physics, Technical University Braunschweig,
D-38106 Braunschweig, Germany}

\author{Wolfram Brenig}

\email{w.brenig@tu-bs.de}

\selectlanguage{english}%

\affiliation{Institute for Theoretical Physics, Technical University Braunschweig,
D-38106 Braunschweig, Germany}

\date{\today}
\begin{abstract}
We study the dynamical thermal conductivity of the Kitaev spin model on a two-leg
ladder. In contrast to conventional integrable one-dimensional spin systems, we show
that heat transport is completely dissipative.  This is a direct consequence of
fractionalization of spins into mobile Majorana matter and a static $Z_{2}$ gauge
field, which acts as an emergent thermally activated disorder. Our finding rests on
three complementary calculations of the current correlation function, comprising a
phenomenological mean-field treatment of thermal gauge fluctuations, a complete
summation over all gauge sectors, as well as exact diagonalization of the original
spin model. The results will also be contrasted against the conductivity discarding
gauge fluctuations.
\end{abstract}

\pacs{75.10.Jm, % Heisenberg model, quantized spin models, quantum spin frustration,
 % magnetic ordering (quantized spin model)
75.10.Pq, , % Spin Hamiltonians
75.10.Kt, % quantum spin liquids, valence bond phases
}

\maketitle
Thermal transport by magnetic degrees of freedom in insulating quantum magnets is a
fascinating probe of spin dynamics. This includes conventional magnons in
two-dimensional (2D) antiferromagnets (AFMs) with long range order (LRO) \cite{Hess03}, but
also more exotic elementary excitations, ranging from gaped triplons in quantum
disordered 1D spin ladder compounds \cite{Sologubenko00,Hess01}, via fractional
spinons in 1D Heisenberg AFM spin chains \cite{Sologubenko00a,Hess2007d,Hlubek2010a}
and potentially also in 2D triangular AFMs \cite{Yamashita2010a}, up to emergent
monopoles in spin ice \cite{Kolland2012a,Toews2013a}.  Recently quantum magnets with
strong spin-orbit coupling have experienced an upsurge of interest, since they
may realize highly frustrated spin systems with directionally dependent
\emph{compass} exchange \cite{Jackeli2009a,Chaloupka2010a,Nussinov2015a}. This
includes Kitaev's model \cite{Kitaev2006}, which represents a rare case of an exactly
solvable, interacting quantum many-body system in 2D, where spin-$1{/}2$ moments on
the honeycomb lattice fractionalize to form an infinite set of spin liquids with
topological degeneracy, comprising Majorana fermions, coupled to a $Z_{2}$ gauge
field \cite{Kitaev2006,Feng2007,Chen2008a,Nussinov2009,Mandal2012a}.  In a broader
context, Kitaev's model is therefore related to topological insulators
\cite{Moore2010a}, superconductors \cite{Alicea2012a}, or fractional quantum Hall
systems \cite{Laughlin1983a}, as well as topological matter and order
\cite{Wen2002a,Wen2006a}.  Interestingly, the physics of Kitaev's model can be
generalized to 3D \cite{Mandal2009,Modic2014a} lattices as well as to 1D
\emph{ladder} versions of the Kitaev model, which allow for topological string order
\cite{Feng2007}.

In this work we shed light on fractionalization as seen by magnetic heat transport in
Kitaev ladders. Theoretical studies of transport in quasi 1D quantum magnets have a
long and fertile history. One key question is the dissipation of currents, which has
been investigated extensively at zero frequency (DC) and momentum in connection with
the linear-response Drude weight (DW)
\cite{heidrichmeisner2007,shastry1990,zotos1997,narozhny1998,zotos1999,kluemper2002,fujimoto2003,heidrichmeisner2003,benz2005,prosen2011,herbrych2011,prosen2013,steinigeweg2013,Steinigeweg2014a,steinigeweg2015}.
The DW is the nondissipating DC part of the current autocorrelation function and, if
existent, indicates a ballistic channel close to equilibrium. In generic
nonintegrable systems, it is unlikely, that DWs exist in the thermodynamic limit
\cite{heidrichmeisner2003}, while the picture is more involved in the integrable case
and depends on the type of current \cite{heidrichmeisner2007}. In general however,
the \emph{energy }conductivity is expected \cite{Jung06} to be infinite, as for the
Heisenberg chain \cite{Zotos97,Klumper02,Zotos2004a}, which implies a finite thermal
DW. This conventional wisdom has recently been confirmed also for Kitaev-Heisenberg
\emph{chains} \cite{Steinigeweg2013a}, which exhibits several integrable points, at
all of which the energy DW is finite. Breaking integrability, e.g., by coupling
Heisenberg chains to form spin ladders routinely renders heat flow dissipative
\cite{Zotos04,Jung06,Karrasch2015a,Steinigeweg2016a}, suppressing the DW. Kitaev
ladders, however, remain an integrable, translationally invariant 1D spin system
suggesting infinite heat conductivity.  Instead of this, and as a prime result of
this work, we show that heat transport is completely \emph{dissipative}. This is a
direct consequence of the matter-gauge-field interactions and can be viewed as a
fingerprint of this $Z_{2}$ spin liquid in 1D.  This behavior is also in sharp
contrast to that of the Kitaev chain \cite{Steinigeweg2013a}, which hosts no gauge
field.  To justify these claims, we calculate the
dynamic energy current correlation function (i) analytically discarding gauge field
excitations, (ii) contrast this against numerical evaluations on large systems
allowing for gauge field excitations in a phenomenological mean-field
approximation, as well as on smaller systems by (iii) complete summation over all
gauge sectors, and (iv) exact diagonalization (ED) of the original spin model.

\begin{figure}[tb]
\begin{centering}
\includegraphics[width=0.65\columnwidth]{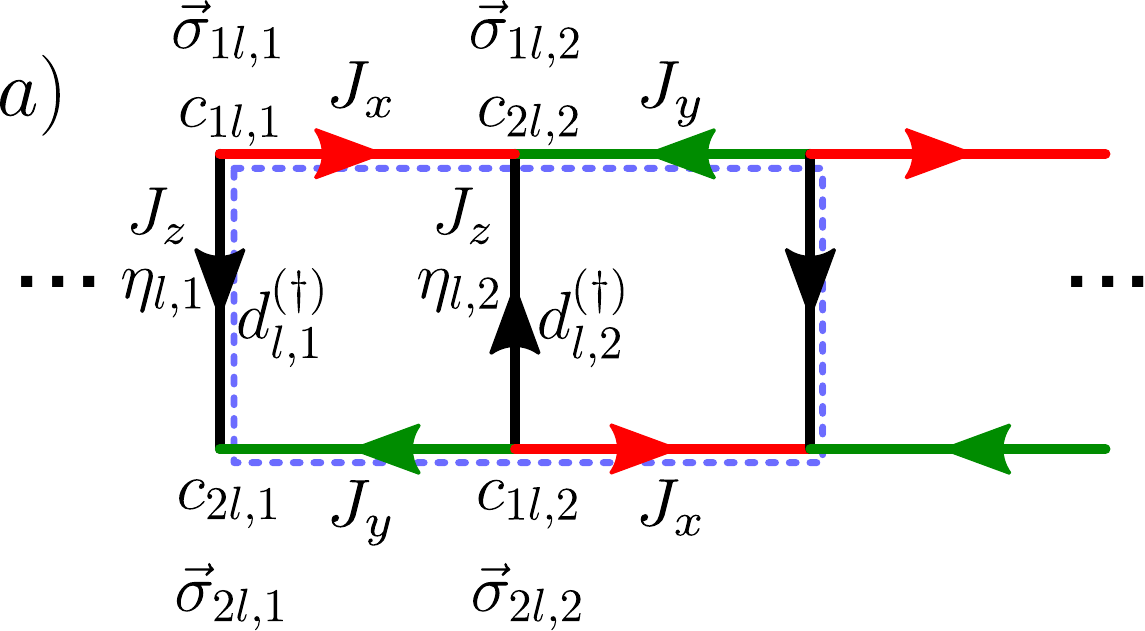}
\par\end{centering}
\begin{centering}
\vspace{2mm}
\par\end{centering}
\centering{}\includegraphics[width=0.65\columnwidth]{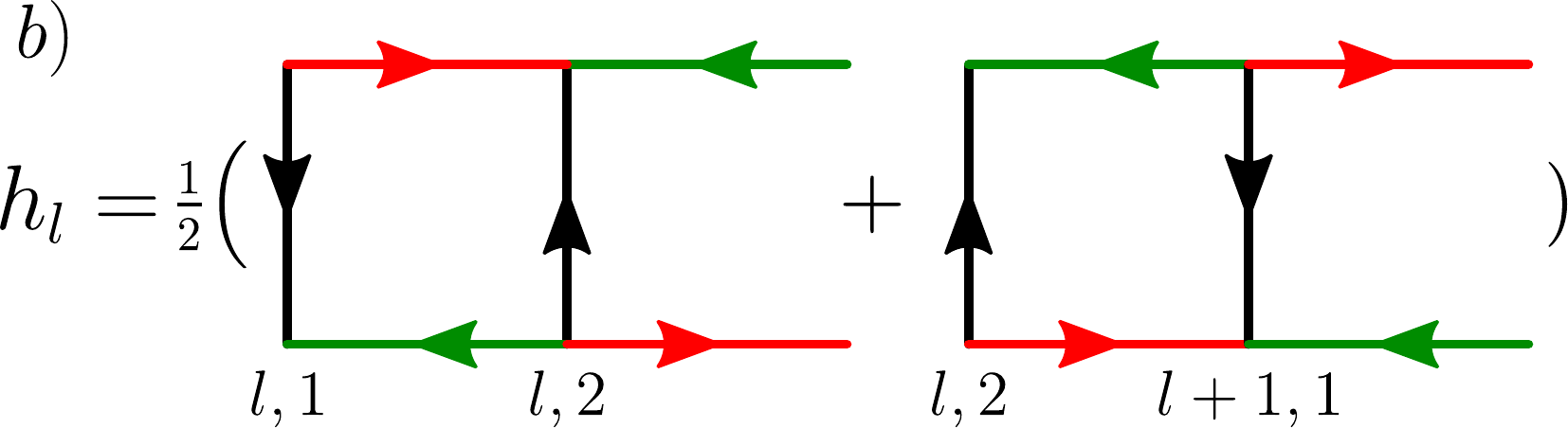}
\caption{\label{fig1}a) Kitaev ladder. Three equivalent representations: spins,
Majorana fermions, and spinless fermions. $J_{x{,}y{,}z}$ exchange interaction. Index
$l$ refers to unit cell (PBC).  $\vec{\sigma}_{il,j}$ Pauli matrix on leg(rung)
$i(j){=}1{,}2$. $c_{il,j}$ Majorana fermion of type(on rung)
$i(j){=}1{,}2$. $d_{l{,}j}^{(\dagger)}$ spinless fermion on rung $j$. Spins and
Majorana fermions located at the vertices, spinless fermions at the center of the
rungs. $\eta_{l{,}j}{=}\pm1$ static $Z_{2}$ gauge fields.  Arrows denote ordering of
Majorana fermions on a bond \cite{Nussinov2015a}. Dashed blue loop of six sites refers
to conserved flux operator $\Phi$ b) Local energy density used in this work.}
\end{figure}

The Hamiltonian of the Kitaev ladder reads 
\begin{equation}
H=\sum_{\langle m,n\rangle}J_{\alpha}\sigma_{m}^{\alpha}\sigma_{n}^{\alpha}
\,\,,\label{eq:1}
\end{equation}
with notations detailed in Fig. \ref{fig1}. It is known that this model can be
mapped onto free Majorana fermions in the presence of static $Z_{2}$ gauge
fluxes. The allowed values $\pm1$ of the latter correspond to the eigenvalues of the
conserved operator $\Phi=\prod_{i=1\dots6}\sigma_{l}^{\alpha(l)}$ around each
six-site loop, indicated in Fig.~\ref{fig1}, where $\alpha(l)=x,y,z$,
refers to that component of the exchange link which is \emph{not} part of loop
passing site $l$ \cite{Kitaev2006,Nussinov2015a}. This mapping can be achieved along
different routes, e.g. using an overcomplete set of four Majorana fermions per site
\cite{Kitaev2006}, the Jordan Wigner transformation \cite{Feng2007}, or bond
algebras \cite{Nussinov2009,Nussinov2015a}.  Following the latter, each exchange link
$\sigma_{m}^{\alpha}\sigma_{n}^{\alpha}$ in Fig. \ref{fig1} is replaced by
$i\eta_{mn}c_{m}c_{n}$, with Majorana fermions $c_{m(n)}$, with $c_{m}^{2}=1$ and
$\{c_{m},c_{n}\}=2\delta_{mn}$, and a \emph{static} $Z_{2}$ gauge field
$\eta_{mn}=\pm1$. The ordering of the Majorana fermions is encoded in the arrows in
Fig. \ref{fig1}, namely within $\eta_{mn}c_{m}c_{n}$ the right fermion refers to the
tip of the arrow. As shown in refs. \cite{Nussinov2009,Nussinov2015a}, the complete
Hilbert space of (\ref{eq:1}) is accounted for by constraining $\eta_{mn}\equiv1$
along the $J_{x,y}$-legs, i.e.
\begin{align}
H= & i\sum_{l}[\eta_{l,1}c_{1l,1}c_{2l,1}+\eta_{l,2}c_{1l,2}c_{2l,2}+j_{x}(c_{1l,1}c_{2l,2}+\nonumber \\
 & c_{1l,2}c_{2l+1,1})+j_{y}(c_{1l,2}c_{2l,1}+c_{1l+1,1}c_{2l,2})]\,,\label{eq:2}
\end{align}
with $j_{x,y}=J_{x,y}/J_{z}$ and $J_{z}=1$. Each \emph{pair} of Majorana fermions
$c_{1(2)l,j}$, can be replaced by \emph{one} spinless fermion, $d_{l,j}^{(\dagger)}$,
using $c_{1l,j}=d_{l,j}^{\dagger}+d_{l,j}^{\phantom{\dagger}}$,
$c_{2l,j}=i(d_{l,j}^{\dagger}-d_{l,j}^{\phantom{\dagger}})$, mapping
Eqn. (\ref{eq:2}) to a BCS Hamiltonian with a two-site basis. Therefore, for $N$ unit
cells, the initial $2^{4N}$ dimensional Hilbert space of Eqn. (\ref{eq:1}) decomposes
into $2^{2N}$ matter sectors, each of which comprises $2^{2N}$ spinless fermion
degrees of freedom. For periodic boundary conditions (PBC), each eigenstate of
(\ref{eq:2}) is (at least) two-fold degenerate by changing sign of all $\eta_{l,j}$.

In the thermodynamic limit, the ground state of Eqn. (\ref{eq:2}) is obtained from a
choice of $\eta_{l,j}\equiv\eta_{j}$, which is translationally invariant with respect
to the unit cell \cite{Lieb1994a,Feng2007,Wu2012a}.  For the parametrization of
Eqn. (\ref{eq:2}) this is $\eta_{1}=-\eta_{2}=\pm$1 \cite{GaugeConvention}.  After
Fourier and Bogoliubov transformation to new spinless fermion quasiparticles
$a_{k,i}^{(\dagger)}$ the Hamiltonian reads
\begin{equation}
H=\sum_{k,i}\epsilon_{k,i}(2a_{k,i}^{\dagger}a_{k,i}^{\phantom{\dagger}}-1)
\label{eq:5}
\end{equation}
with energies $2\epsilon_{k,1(2)}=2[j_{+}^{2}c^{2}+(1\varpm j_{-}s)^{2}]^{-1/2}$, with
$c=\cos(k/2)$, $s=\sin(k/2)$, $j_\pm = j_x\pm j_y$, and the Brillouin zone fixed to
$k\in[0,2\pi[$ \cite{Feng2007,Feng2007Comment,NoteOnIsing}.  For
convenience we redefine the quasiparticles within the extended zone scheme
$k\not\in[0,2\pi[$ to satisfy $\epsilon_{k,1(2)}=\epsilon_{-k,1(2)}$.

The main goal of this paper is to analyze the finite temperature energy current
correlation function $C(t)=\langle J(t)J\rangle/N$ and its Fourier transform
$C(\omega)=\int_{-\infty}^{\infty}dt\,C(t)\exp(i\omega t)=
C_0\delta(\omega)+C(\omega\neq0)$, encoding the physics of the thermal
conductivity \cite{steinigeweg2015}.  Here, $\langle\dots\rangle$ is the canonical
thermal trace at temperature $T=1/\beta$ ($k_B=1$). The energy current $J$ follows
from the energy polarization $P=\sum_{l}l\,h_{l}$ through $J=i[H,P]$ ($\hbar=1$),
where $h_l$ is the energy density depicted in Fig. \ref{fig1}b).  $D\equiv C_0/T^2$
is the Drude weight \cite{NoteOnDefD}, which quantifies the non-dissipative current
dynamics. Since the energy current is diagonal in the gauge fields, one may write
\begin{equation}
\langle J(t)J\rangle=
Tr_{\eta}[Z_{d(\eta)}\,\langle J(t)J\rangle_{d(\eta)}]/Z\,. \label{eq:4}
\end{equation}
The subscript $d(\eta)$ refers to tracing over matter fermions only at a \emph{fixed}
gauge field state. The trace over gauge field states $Tr_{\eta}$ can be treated in
different ways \cite{Motome2014a,Motome2015b,Motome2015a,NoteOnMotome}. Here we
consider two approaches: (i) we perform exact summation over all gauge sectors on
small systems to compare with ED of the original spin model, and (ii) we approximate
$Tr_{\eta}$ in a mean-field sense.

For the latter, we envisage tracing over $\eta$ by joining ground state gauge domains
of either $\eta_1=-\eta_2=+1$ or $-1$ of arbitrary lengths $L$.
At $T=0$, forming a single domain in the ground state is gaped by
$\Delta_L$, referring to the cost of creating two gauge fluxes at the domain
walls \cite{Kitaev2006}. The domain walls are deconfined, with
$\Delta_{L\rightarrow\infty}$ converging to some constant \cite{DofL}. To simplify, we
set $\Delta_L\approx\Delta$, with $\Delta$ evaluated by flipping a single
$\eta_{l,i}$. Finally we approximate the preceding to remain true at $T\neq 0$,
i.e. we discard thermal polarization effects on $\Delta$ from the matter fermions. This maps the
gauge field thermodynamics to that of the $S{=}\pm 1$ nn-Ising chain with exchange
constant $\Delta/4$. To perform the $Tr_{\eta}$ we then confine the summation to
gauge field states, which only contain a temperature dependent \emph{mean}, even
number of domain walls $n(T)$. We fix $n(T)$ by using that the
average nn spin correlation function $c_{1}=\sum_{l=1}^{2N}\langle S_l
S_{l+1}\rangle$, which is known for the 1D Ising model \cite{Ising1925}, satisfies
$c_1=2N-2n(T)$, yielding
\begin{equation}
n(T) = 2 N /(e^{\Delta/2T} +1)\, ,
\label{eq:3}
\end{equation}
rounded to multiples of two. With this Eq. (\ref{eq:4}) reads
\begin{equation}
\langle J(t)J\rangle\approx\langle\langle J(t)J\rangle_{d(\eta)}\rangle_{n(T)}
\,,\label{eq:7}
\end{equation}
where $\langle\ldots\rangle_{n(T)}$ refers to random averaging over gauge domains
with a number of walls set by $n(T)$.

In turn, evaluating $\langle J(t)J\rangle$ reduces to a \emph{disorder problem} with
a temperature induced 'defect' density. We emphasize, that neither the neglect of
fluctuations in $n(T)$, nor its specific dependence on $T$ is qualitatively relevant
for our main conclusions, as long as $n(T)$ smoothly interpolates between an
exponential on-turn and a random state at $T=\infty$.  Furthermore, our approach
manifestly ensures that the ladder shows no LRO in the gauge field at any $T\neq 0$,
since for $n(T)\neq 0$ domains of arbitrary size and location are included in the
trace.

To appreciate the impact of the thermal fluctuations of the gauge field on the
transport, we first suppress the trace over $\eta_{l,i}$, and pick only the
\emph{clean} ground state gauge, allowing however for finite temperatures. Using the
energy density of Fig. \ref{fig1}b), expressed in terms of the \emph{original} matter
fermions $d_{lj}^{(\dagger)}$, deriving the current, and after transforming to
Bogoliubov particles, one gets
\begin{eqnarray}\label{eq:6}
J &=& \sum_{k, i} u_{k,i} (a_{k,i}^\dag a_{k,i}+a_{-k,i}a_{-k,i}^\dag)
\nonumber \\ 
&+&  j_{k,i} (a_{k,i}^\dag a_{-k,i}^\dag + a_{-k,i}a_{k,i}) 
\end{eqnarray} 
with $i=1,2$, $u_{k,i}=(j_{+}^{2}-j_{-}^{2})\sin(k)/2+(-1)^i j_{-}\cos(k/2)$, and
$j_{k,i}=(-1)^i|j_{+}\cos(k/2)|$.  Using (\ref{eq:5}) and (\ref{eq:6}), solving for
$C(t)$ is straightforward. For the Fourier transform $C(\omega)$ we obtain
\begin{eqnarray}
C(\omega) &=& \frac{4\pi}{N}\sum_{k,i} \left\{
2u_{k,i}^2 f_{k,i}(1-f_{k,i}) \delta(\omega) + j_{k,i}^{2}
\left[f^2_{k,i} \right. \right. \nonumber \\
& & \left.\left. \times\delta(\omega + 4\epsilon_{k,i}) 
+(1-f_{k,i})^2\delta(\omega-4\epsilon_{k,i})
\right]\right\},\hphantom{aa} 
\label{eq:8}
\end{eqnarray} 
with $f$ being the Fermi distribution, $f_{k,i} = 1/(e^{2\beta \epsilon_{k,i}}+1)$.
This result is of the form typical for a clean superconductor, comprising a zero
frequency quasiparticle Drude peak and two finite frequency pair breaking spectra,
corresponding to the two quasiparticle energies of Eqn. (\ref{eq:5}). 

\begin{figure}[tb]
\begin{centering}
\includegraphics[width=0.7\columnwidth]{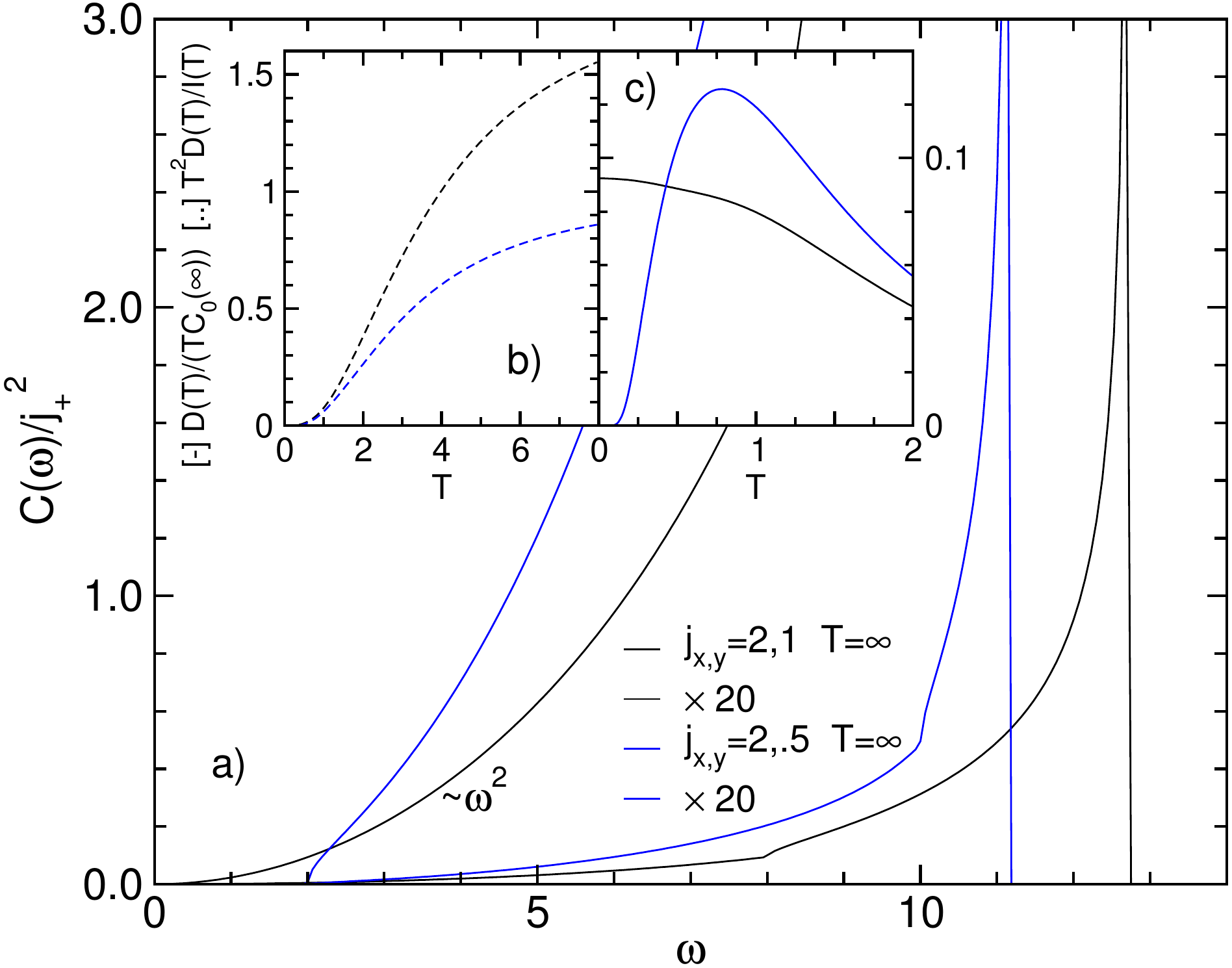}
\par\end{centering}
\caption{\label{fig:3}Black(blue) lines: infinite temperature dynamical current
correlation function $C(\omega)$ versus frequency $\omega>0$ using
the ground state gauge for gapless(ful) matter sector at $j_{x,y}=2,1(2,.5)$.
Inset: DW $D(T)$ versus temperature $T$ normalized
to $I(T)/T^2\,${[}$T C_0(T=\infty)${]} dashed{[}solid{]}.}
\end{figure}

In Fig.~\ref{fig:3} the current correlation function is shown for two representative
cases of $j_{x,y}$, referring to a gapless (gaped) matter sector at $j_{x,y}=2,1$
($j_{x,y}=2,0.5$).  Several comments are in order. First, the regular spectrum
$C(\omega\neq0)$ is depicted only for $\omega>0$, since
$C(-\omega)=e^{-\beta\omega}C(\omega)$, as required by detailed balance. Second, in
the gapless case the regular spectrum for $\omega\ll 1$ shows a power law
$C(\omega)\propto\omega^{2}$ due to $j_{\pi+q,i}^2\propto q^2$, while displaying a
van-Hove singular gap for $|j_{-}|\neq1$.  No qualitative difference arises in
$C(\omega)$ between the topologically trivial and nontrivial phases, as to be
expected for the current of a local energy density.  At elevated energies two more
van-Hove singularities arise, one at the onset of the the second quasiparticle
excitations and one at the upper band edge.  The insets Fig. \ref{fig:3}b) and c)
detail the Drude weight versus temperature, relative to its integrated regular spectral weight
$I(T)=-\hspace{-9pt}\int_{-\infty}^{\infty}d\text{\ensuremath{\omega}}\,C(\omega)$,
skipping the Drude peak, and relative to the high temperature value.  
Fig. \ref{fig:3}b) shows that $D(T)$ is finite for any 
$T\neq0$ and that $T^2 D(T)$ is comparable to $I(T)$ at sufficiently large
temperatures.  Fig. \ref{fig:3}c) proves that $D(T\ll1)\propto T$ as is true for 1D free
fermions irrespective of the actual dispersion. In the gaped case $D(T)$ is
exponentially activated.

Now we include the gauge fluctuations via Eqn. (\ref{eq:7}). This requires a
numerical treatment. We define a $4N$ component operator
$\mathbf{D}^{\dagger} = (d_{1,1}^{\dagger}, d_{1,2}^{\dagger}\ldots 
d_{N,1}^{\dagger},d_{N,2}^{\dagger},d_{1,1}^{\phantom{\dagger}},$
$d_{1,2}^{\phantom{\dagger}}\ldots d_{N,1}^{\phantom{\dagger}},
d_{N,2}^{\phantom{\dagger}})$ of the original matter fermions, in terms of which the
Hamiltonian and the current are set up in real space as $H=\mathbf{D}^{\dagger}
\mathbf{h} (\eta) \mathbf{D}$ and
$J=\mathbf{D}^{\dagger}\mathbf{j}(\eta)\mathbf{D}$.  Both,
$\mathbf{h}(\eta)$ and $\mathbf{j} (\eta)$ are $4N\times4N$ matrices, which depend on
the actual state of the gauge field $\eta=\eta_{11}, \eta_{22} \ldots
\eta_{N1}, \eta_{N2}$.  For each given $\eta$ we compute a Bogoliubov
transformation $\mathbf{U}$, which introduces canonical quasiparticle fermions
$\mathbf{A}^{\dagger} = (a_{1}^{\dagger},\ldots
a_{2N}^{\dagger},a_{1}^{\phantom{\dagger}}, \ldots a_{2N}^{\phantom{\dagger}})$ via
$\mathbf{A} = \mathbf{U}^{\dagger} \mathbf{D}$ and maps the Hamiltonian to
$H=\mathbf{A}^{\dagger} \mathbf{E}\mathbf{A}$, where $\mathbf{E}$ is diagonal and
$\mathrm{diag}(E)=(\varepsilon_{1}\ldots \varepsilon_{2N},
-\varepsilon_{1}\ldots-\varepsilon_{2N})$ are the quasiparticle
energies. With these definitions the current correlation function reads
\begin{align} 
C(\omega)= & \frac{2\pi}{N}\sum_{\kappa\lambda\mu\nu}L_{\kappa\lambda}
L_{\mu\nu}(\langle A_{\kappa}^{\dagger}A_{\nu}^{\phantom{\dagger}}
\rangle\langle A_{\lambda}^{\phantom{\dagger}}A_{\mu}^{\dagger}\rangle
\nonumber \\
 & -\langle A_{\kappa}^{\dagger}A_{\mu}^{\dagger}\rangle\langle 
A_{\lambda}^{\phantom{\dagger}}A_{\nu}^{\phantom{\dagger}}\rangle)
\delta(\omega-2(\varepsilon_{\kappa}-\varepsilon_{\lambda}))\,,\label{eq:9} 
\end{align}
where $\mathbf{L}=\mathbf{U}^{\dagger}\mathbf{j}(\eta)\mathbf{U}$ and $\langle
A_{\mu}^{(\dagger)}A_{\nu}^{(\dagger)}\rangle$ is either zero, $f_\mu$,  or 
$(1-f_\mu)$, depending on the components of the spinor $\mathbf{A}$ involved.

\begin{figure}[tb]
\begin{centering}
\includegraphics[width=0.8\columnwidth]{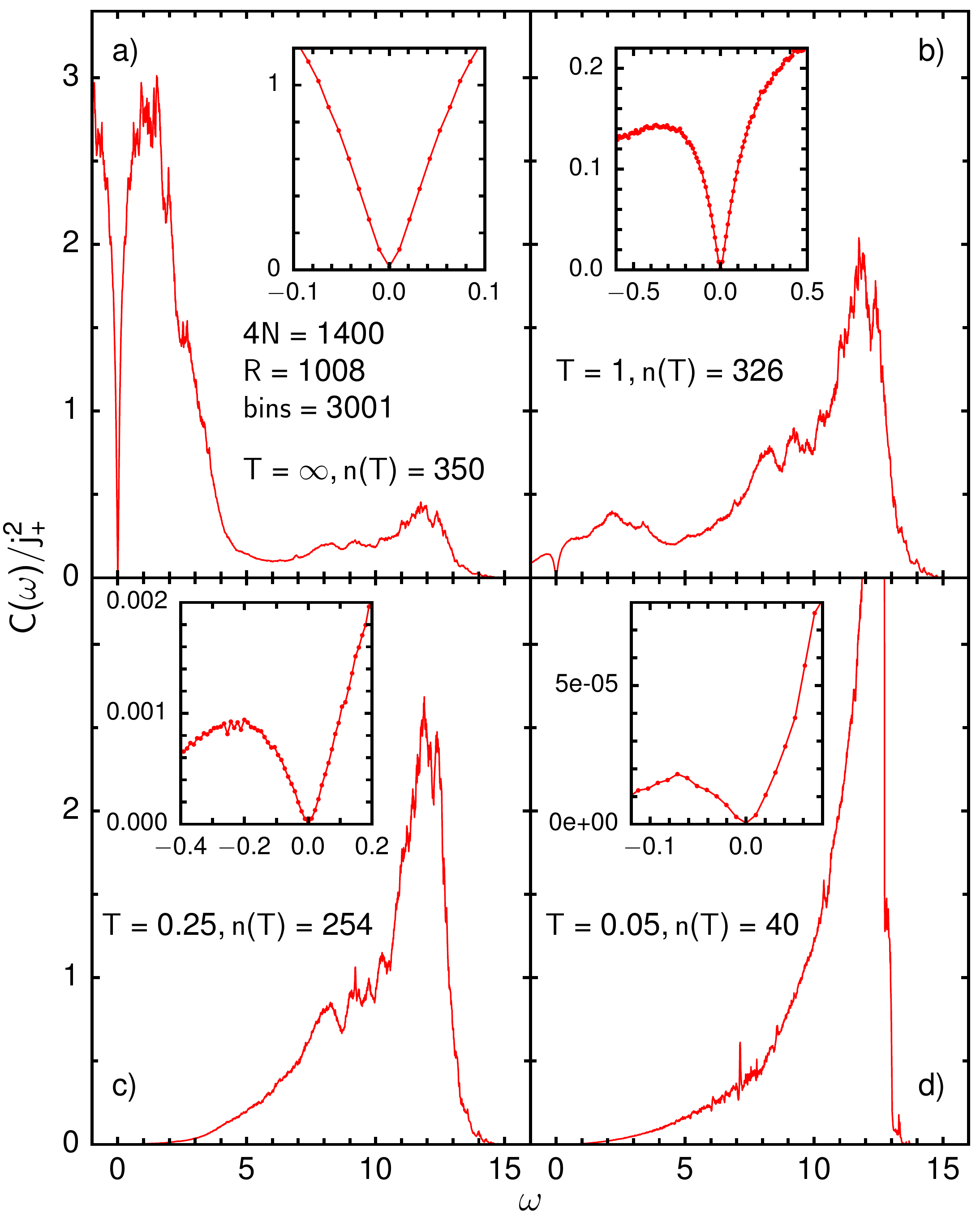}
\par\end{centering}

\caption{\label{fig:4}Current correlation function $C(\omega)$ versus
frequency for $j_{x,y}{=}2,1$ and various temperatures $T=\infty,\ldots,0.05$, from a)
to d) corresponding to gauge domain wall numbers $n(T)= N,\ldots,\ll N$.
$\Delta(j_{x,y}{=}2,1)\approx 0.279$. Negative frequencies included to
highlight mobility gap. Inset: low-$\omega$ behavior.}
\end{figure}

Fig.~\ref{fig:4} shows results for $C(\omega)$ from Eq.~(\ref{eq:9}), for
$j_{x,y}=2,1$ on lattices with 1400 sites, by binning the $\delta$-functions in
windows of the order $10^{-2}$.  We perform an average over 1008 random gauge domain
configurations, with wall numbers (\ref{eq:3}), for various temperatures, chosen for
$n(T)$ to span a typical set of concentrations of domain walls, ranging from almost
the clean limit to the maximum possible number of domains. These results are in stark
contrast to those of Fig.~\ref{fig:3}. First, the Drude weight is zero at any
temperature. Instead, at high temperatures, where the Drude weight in the clean limit
is a substantial fraction of the total integrated weight, Fig. \ref{fig:4}a) can
rather be interpreted as the Drude weight being shifted into a range of finite low
frequencies, by scattering from the gauge excitations. Similar physics, albeit weaker
is visible also in Fig. \ref{fig:4}b).  Each inset in Fig. \ref{fig:4} details
$C(|\omega|\ll1)$, clearly evidencing a mobility gap with a vanishing DC correlation
function $C(\omega\rightarrow0)=0$.  We emphasize, that the energy scale of the
mobility gap is unrelated to that of the $\omega^2$ power law of gapless case in
Fig. \ref{fig:3}. Rephrasing our results: heat \emph{localizes} because the Kitaev
ladder comprises 1D free matter fermions scattering off a disordered static gauge
potential \cite{Gof4}. This also clarifies our results to be qualitatively
\emph{insensitive} to details of the form of $n(T)$ and the inclusion of fluctuations
around Eq. (\ref{eq:3}).  The fine structure in $C(\omega)$ comprises effects of
finite size, finite domain realization number, but also scattering from ``typical''
clusters of excited gauge fields, as eg. the clear case of impurity anti-bound
states, visible in Fig.~\ref{fig:4}d) above the bare spectral cut-off. Nota bene,
Eqn. (\ref{eq:8}) implies $C(\omega>0,T=0)=4C(\omega>0,T=\infty)$ and moreover,
because of $n(T\rightarrow 0)=0$, the spectrum from the numerical real space calculation should
approach that of Fig. \ref{fig:3} up to a constant as $T\rightarrow 0$. This agrees with the evolution of
Fig. \ref{fig:4} a) to d).

\begin{figure}[tb]
\begin{centering}
\includegraphics[width=0.71\columnwidth]{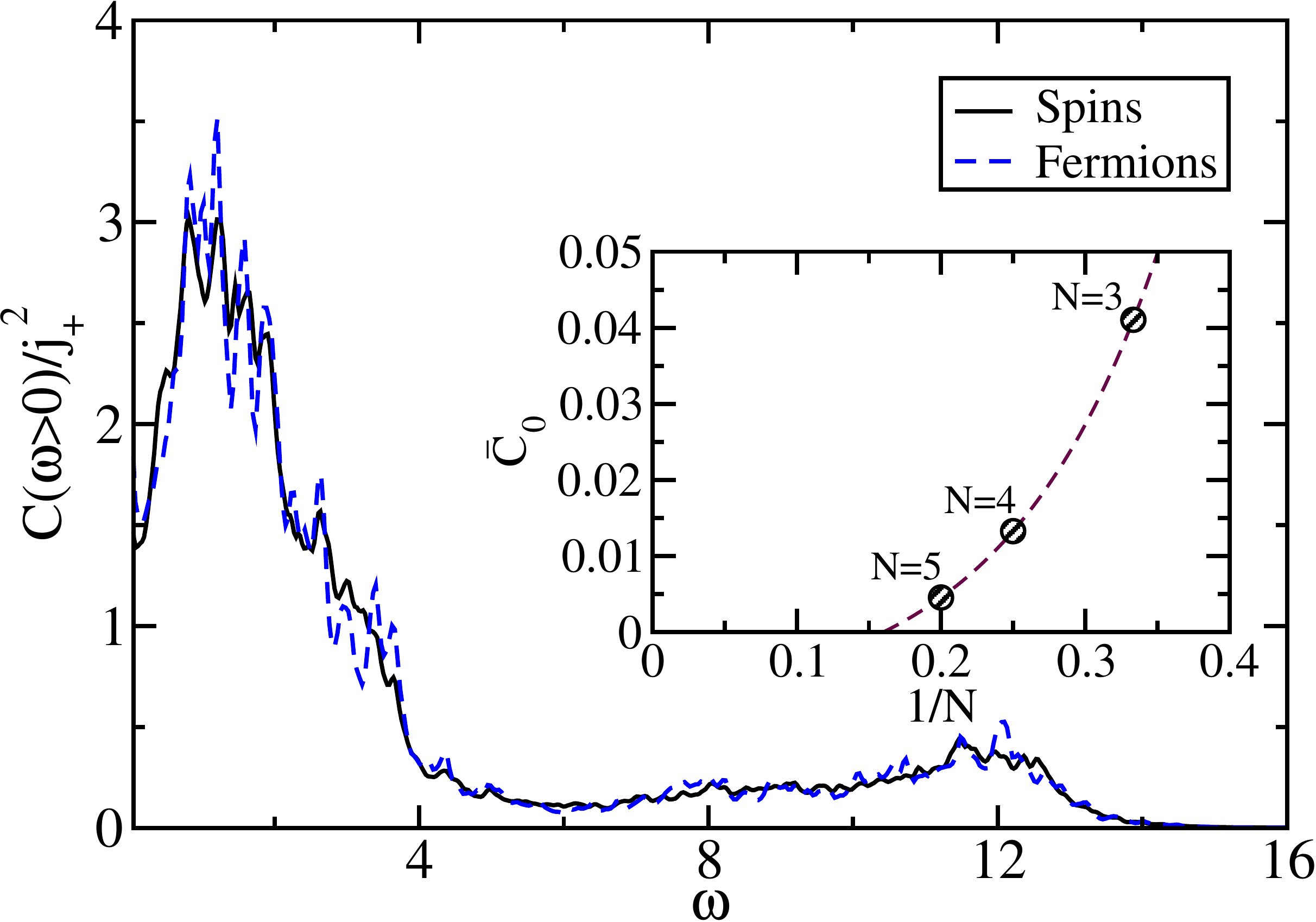}
\par\end{centering}

\caption{\label{fig:5} $C(\omega)$ from ED of the original spin Hamiltonian  (black solid line) 
and from averaging over all sectors of the  fermionic model (blue dashed line). 
Inset: Finite size scaling of the DW, evaluated in the spin model.}   
\end{figure}

To further substantiate our results, we also perform numerically exact evaluation of
$C(\omega)$ in the full many body Hilbert space of the original spin model, to
compare it with an evaluation in the full Hilbert space of the fermionic model. For
the spin model, the canonical average of Eq.~\eqref{eq:4} is carried out on the
Hamiltonian basis, obtained via ED, on systems up to 20 spins. For the purpose of
this calculation, $\delta$-functions are approximated by Lorentzians with a half
width parameter of the order of $10^{-2}$. Fig. \ref{fig:5} shows results for
$\beta=0$. First, the agreement of the two calculations is impressive. The
differences are due to the neglect of boundary terms \cite{Mandal2012a} in mapping
Eq. \eqref{eq:1} to \eqref{eq:2}. Second, these results corroborate our findings from
the disorder averaging scheme, with $C(\omega)$ being in good qualitative agreement
with the high temperature results presented in Fig.~\ref{fig:4}a).  Note that the
seemingly finite value of $C(\omega\rightarrow0)$ is an artifact of the Lorentzian
broadening of the $\delta$-functions combined with the steep dip of $C(|\omega|\ll
1)$, observable in Fig.~\ref{fig:4}a). This cannot be captured on small systems.
Inevitably for finite systems, we also find a finite DW, which however, scales to
zero at least exponentially in the thermodynamic limit. This is shown in the inset of
Fig.~\ref{fig:5}, where we present $C_0$ divided by the sum rule, i.e. $\bar C_0 = N
C_0/\langle J J \rangle$, as a function of the inverse system size. This is in stark
contrast to ED calculation for the DW of the Kitaev \emph{chain}
\cite{Steinigeweg2013a}, for which the DW is essentially independent of system size
and finite.

In conclusion we have shown that, even though pure Kitaev ladders are translationally
invariant and integrable spin systems, they are perfect heat insulators due to
fractionalization of spins into mobile Majorana matter and static $Z_{2}$ gauge
fields, which generate an emergent disorder at finite temperature. This is different
from Kitaev chains. In Kitaev models with $d\geq 2$, thermal currents will scatter
off thermally excited $Z_{2}$-links similarly, connecting this physics to transport
in higher dimensional superconductors with a temperature dependent impurity
concentration. Non-Kitaev exchange will lead to dispersion of gauge excitations,
restoring translational invariance below some energy scale, where Majorana particle
heat transport may dissipate by relaxing momentum into mobile gauge excitations.

\emph{Acknowledgments.} We thank R. Steinigeweg, M. Vojta, S. Rachel,
and J. van den Brink for fruitful discussions and comments. Work of
W.B. has been supported in part by the DFG through SFB 1143 and the
NSF under Grant No. NSF PHY11-25915. W.B. also acknowledges kind hospitality
of the PSM, Dresden. 

\bibliographystyle{apsrev}

\begin{thebibliography}{55}
\bibitem{GaugeConvention}
To arrive at a notation more reminiscent of that traditionally used for the 2D case,
one may relabel $\eta_{l,2}\rightarrow-\eta_{l,2}$, to obtain a completely
'homogeneous' gauge.
\bibitem[{Not({\natexlab{a}})}]{NoteOnIsing}
\bibinfo{note}{At $j_{x,y}=0$ this is consistent with an excitation energy of 2
  per Ising z-link of type $\sigma^z_1\sigma^z_2$}.
\bibitem{DofL}
Convergence of $\Delta_L$ to $\Delta_\infty$, will be exponential for any
non-critical, i.e. gaped matter fermion case, i.e. $|j_-|\neq 1$, while at $|j_-|=1$
algebraic behavior might arise.
\end{thebibliography}

\end{document}